\documentclass[10pt,a4paper,final]{iopart}
\usepackage{fancyhdr}
\usepackage{iopams}
\usepackage{graphicx}
\usepackage[breaklinks=true,colorlinks=true,linkcolor=magenta,urlcolor=blue,citecolor=red]{hyperref}
\usepackage{aas_macros}
\newcommand{\pa}{\partial}
\newcommand{\nn}{\nonumber}

\newcommand{\be}{\begin{equation}}
\newcommand{\ee}{\end{equation}}
\newcommand{\vp}{\varphi}

\newcommand{\defeq}{\equiv}
\newcommand{\M}{{\rm{M}}}

\newcommand{\tc}{\rm{c}}
\newcommand{\ts}{\rm{s}}

\newcommand{\Mb}{\tilde{M}_{\rm{b}}}

\newcommand{\pJ}{\tilde{p}}
\newcommand{\qJ}{\tilde{q}}

\newcommand{\omegab}{\bar{\omega}}

\newcommand{\uJ}{\tilde{u}}

\newcommand{\rhoJ}{\tilde{\epsilon}}

\newcommand{\sigmaJ}{\til{\sigma}}
\newcommand{\gJ}{\tilde{g}}

\newcommand{\OmegaJ}{\Omega}

\newcommand{\RE}{R_{\ast}}
\newcommand{\til}{\tilde}
\newcommand{\eqref}[1]{(\ref{#1})}

\begin{document}

\pagestyle{fancy}
\lhead{Slowly Rotating Anisotropic Neutron Stars}
\chead{}
\rhead{\thepage}
\lfoot{}
\cfoot{}
\rfoot{}

\title{Slowly Rotating Anisotropic Neutron Stars in General Relativity and Scalar-Tensor Theory}

\author{Hector O. Silva$^1$, Caio F. B. Macedo$^2$, Emanuele Berti$^{1,3}$, Lu\'is C. B. Crispino$^2$}
\address{
$^1$Department of Physics and Astronomy, The University of Mississippi, University, MS 38677, USA \\
$^2$Faculdade de F\'isica, Universidade Federal do Par\'a, 66075-110, Bel\'em, Par\'a, Brazil \\
$^3$CENTRA, Departamento de F\'isica, Instituto Superior T\'ecnico, Universidade de Lisboa, Avenida Rovisco Pais 1, 1049 Lisboa, Portugal}
\ead{hosilva@phy.olemiss.edu, caiomacedo@ufpa.br, crispino@ufpa.br, eberti@olemiss.edu}

\begin{abstract}
  Some models (such as the Skyrme model, a low-energy effective field
  theory for QCD) suggest that the high-density matter prevailing in
  neutron star interiors may be significantly anisotropic. Anisotropy
  is known to affect the bulk properties of nonrotating neutron stars
  in General Relativity. In this paper we study the effects of
  anisotropy on slowly rotating stars in General Relativity. We also
  consider one of the most popular extensions of Einstein's theory,
  namely scalar-tensor theories allowing for spontaneous scalarization
  (a phase transition similar to spontaneous magnetization in
  ferromagnetic materials). Anisotropy affects the moment of inertia
  of neutron stars (a quantity that could potentially be measured in
  binary pulsar systems) in both theories. We find that the effects of
  scalarization increase (decrease) when the tangential pressure is
  bigger (smaller) than the radial pressure, and we present a simple
  criterion to determine the onset of scalarization by linearizing the
  scalar-field equation.  Our calculations suggest that binary pulsar
  observations may constrain the degree of anisotropy or even, more
  optimistically, provide evidence for anisotropy in neutron star
  cores.
\end{abstract}


\section{Introduction}
\label{sec:intro}

Most investigations of the structure of neutron stars (NSs) assume
isotropic matter with a perfect-fluid equation of state (EoS) relating
the pressure and density in the stellar interior. However, various
physical effects can lead to local anisotropies (see
\cite{HerreraSantos:1997} for a review).
Anisotropy can occur for stars with a solid core
\cite{kippenhahn1990stellar} or strong magnetic fields
\cite{Yazadjiev:2012,Folomeev:2015aua,Kamiab:2015uda}.
Spaghetti- and lasagna-like structures would induce anisotropic
elastic properties that could be important for NS quakes
\cite{Heiselberg:1999mq}.
Nuclear matter may be anisotropic at very high densities
\cite{Ruderman:1972aj,Canuto:1974gi}, where the nuclear interactions
must be treated relativistically and phase transitions (e.g. to pion
condensates \cite{Sawyer:1972cq} or to a superfluid state
\cite{Carter:1998rn}) may occur.
For example, Nelmes and Piette \cite{Nelmes:2012uf} recently
considered NS structure within the Skyrme model, a low energy,
effective field theory for Quantum Chromodynamics (QCD), finding
significant anisotropic strains for stars with mass
$M\gtrsim 1.5M_\odot$ (see also \cite{Adam:2014dqa,Adam:2015lpa}).
From a mathematical point of view, two-fluid systems can be shown to
be equivalent to a single anisotropic fluid \cite{Letelier:1980}.
Anisotropy affects the bulk observable properties of NSs, such as the
mass-radius relation and the surface redshift \cite{BowersLiang:1974}:
it can increase the maximum NS mass for a given EoS
\cite{BowersLiang:1974,Horvat:2010xf} and stabilize otherwise unstable
stellar configurations \cite{Dev:2003qd}.
Incidentally, exotic objects such as gravastars \cite{Cattoen:2005he}
and boson stars \cite{Schunck:2003kk,Macedo:2013jja} are also
equivalent to anisotropic fluids (i.e., they have anisotropic
pressure).

It is known that rotation can induce anisotropy in the pressure due to
anisotropic velocity distributions in low-density systems
\cite{HerreraSantos:1997}, but to the best of our knowledge -- with
the exception of some work by Bayin \cite{Bayin:1982vw} -- slowly
rotating anisotropic stars have never been investigated in General
Relativity (GR). The
goal of this paper is to fill this gap using two different
phenomenological models for anisotropy
\cite{BowersLiang:1974,Horvat:2010xf}, and to extend the analysis of
slowly rotating anisotropic stars to scalar-tensor theories of
gravity.

Scalar-tensor theories are among the simplest and best studied
extensions of GR \cite{Fujii:2003pa}. In addition
to the metric, in these theories gravity is also mediated by a scalar
field. Scalar-tensor theories arise naturally from the dimensional
reduction of higher-dimensional proposals to unify gravity with the
Standard Model, and they encompass $f(R)$ theories of gravity as
special cases \cite{Sotiriou:2008rp,DeFelice:2010aj}.
The simplest variant of scalar-tensor theory, Brans-Dicke theory,
is tightly constrained
experimentally \cite{lrr-2006-3}, but certain versions of the theory
could in principle differ from GR by experimentally measurable amounts
in the strong-field regime, as shown by Damour and Esposito-Far\`ese
\cite{Damour:1992we,Damour:1993hw}.

From an astrophysical standpoint, compact objects such as black holes
and NSs are the most plausible candidates to test strong-field gravity
\cite{Berti:2015itd}. Compared to black holes, NSs are a more
promising strong-field laboratory to distinguish scalar-tensor gravity
from GR, because a large class of scalar-tensor theories admits the
{\em same} black-hole solutions as GR (see \cite{Sotiriou:2011dz} and
references therein), and the dynamics of black holes can differ from
GR only if the black holes are surrounded by exotic forms of matter
\cite{Stefanov:2007eq,Doneva:2010ke,Cardoso:2013opa,Cardoso:2013fwa}
or if the asymptotic behavior of the scalar field is nontrivial
\cite{Horbatsch:2011ye,Berti:2013gfa}.

The study of NS structure in GR is textbook material
\cite{HTWW1965,Misner1973,Shapiro:1983du,FriedmanStergioulas}, and
there is an extensive literature on stellar configurations in
scalar-tensor theories as well (see
e.g.~\cite{Horbatsch:2010hj,Pani:2011xm} and references therein). One
of the most intriguing phenomena in this context is ``spontaneous
scalarization'' \cite{Damour:1993hw}, a phase transition analogous to
the familiar spontaneous magnetization in solid state physics
\cite{Damour:1996ke}: in a certain range of central densities,
asymptotically flat solutions with a nonzero scalar field are possible
and energetically favored with respect to the corresponding GR
solutions.

In the absence of anisotropy, the degree of scalarization depends on a
certain (real) theory parameter $\beta$, defined in Eq.~\eqref{cf}
below. Theory predicts that scalarization cannot occur (in the absence
of anisotropy) when $\beta\gtrsim -4.35$ \cite{Harada:1998ge}.
Present binary pulsar observations yield a rather tight experimental
constraint: $\beta\gtrsim -4.5$ \cite{Freire:2012mg,Wex:2014nva}.
One of our main findings is that the effects of scalarization, as well
as the critical $|\beta|$ for spontaneous scalarization to occur,
increase (decrease) for configurations in which the tangential
pressure is bigger (smaller) than the radial pressure. Therefore
binary pulsars can be used to constrain the degree of anisotropy at
fixed $\beta$, or to constrain $\beta$ for a given degree of
anisotropy. This may open the door to experimental constraints on the
Skyrme model via binary pulsar observations.
Other notable findings of this study are (i) an investigation of the
dependence of the stellar moment of inertia on the degree of
anisotropy $\lambda$ (more precisely, $\lambda_{\rm H}$ and
$\lambda_{\rm BL}$, because we consider two different anisotropy
models \cite{BowersLiang:1974,Horvat:2010xf}); and (ii) an
investigation of the threshold for scalarization for different values
of $\beta$ and $\lambda$ in terms of a simple linear stability
criterion, along the lines of recent work for black holes surrounded
by matter \cite{Cardoso:2013opa,Cardoso:2013fwa}.

The plan of the paper is as follows.
In Section~\ref{sec:STanis} we introduce the equations of motion in
scalar-tensor theory and the stress-energy tensor describing
anisotropic fluids that will be used in the rest of the paper.
In Section~\ref{sec:HT} we present the equations of structure for
relativistic stars at first order in the slow-rotation expansion.
The macroscopic properties of NSs obtained by integrating these
equations for two different models of anisotropic stars are presented
in Section~\ref{sec:results}.
Section~\ref{sec:critical_mass} shows that a linear approximation is
sufficient to identify the threshold for spontaneous scalarization for
different values of $\beta$ and $\lambda$.
Section~\ref{sec:conclusions} summarizes our main conclusions and
points out possible avenues for future work.
Finally, in Appendix \ref{sec:I} we give a detailed derivation of an
integral formula to compute the moment of inertia.
Throughout this work, quantities associated with the Einstein (Jordan)
frame will be labeled with an asterisk (tilde). We use geometrical
units ($c=G_{\ast}=1$) unless stated otherwise and signature $(-,+,+,+)$.

\section{Anisotropic fluids in scalar-tensor theory of gravity}
\label{sec:STanis}

\subsection{Overview of the theory}

We consider a massless scalar-tensor theory described by an
Einstein-frame action \cite{Damour:1993hw,Damour:1996ke}
\begin{eqnarray}
S &= \frac{c^4}{16\pi G_{\ast}}\int d^4x \frac{\sqrt{-g_{\ast}}}{c}
\left( R_{\ast} - 2 g_{\ast}^{\mu\nu}\pa_{\mu}\vp\pa_{\nu}\vp \right)
\nonumber \\
&+ S_\M \left[\psi_\M;A^2(\vp)g_{\ast\mu\nu} \right],
\label{action}
\end{eqnarray}
where $G_{\ast}$ is the bare gravitational constant, $g_{\ast} \defeq
{\rm det}\left[\, g_{\ast\mu\nu}\right]$ is the determinant of the
Einstein-frame metric $g_{\ast\mu\nu}$, $\RE$ is the Ricci curvature
scalar of the metric $g_{\ast\mu\nu}$, and $\varphi$ is a massless
scalar field. $S_\M$ is the action of the matter fields, collectively
represented by $\psi_\M$.  Free particles follow geodesics of the
Jordan-frame metric $\tilde{g}_{\mu\nu}\defeq
A^2(\varphi)g_{\ast\mu\nu}$, where $A(\varphi)$ is a conformal factor.
In this work we assume that $A(\vp)$ has the form
\be
A(\vp) \defeq e^{\frac{1}{2} \beta \vp^2},
\label{cf}
\ee
where $\beta$ is the theory's free parameter and, as we recalled in
the introduction, current binary pulsar observations constrain it to
the range $\beta\gtrsim -4.5$ \cite{Freire:2012mg,Wex:2014nva}.

The field equations of this theory, obtained by varying the action $S$
with respect to $g^{\mu\nu}_{\ast}$ and $\vp$, are given by
\begin{eqnarray}
R_{\ast\mu\nu} &= 2 \pa_{\mu}\vp\pa_{\nu}\vp + 8 \pi \left( T_{\ast\mu\nu} - \frac{1}{2} T_{\ast} g_{\ast\mu\nu}\right),
\label{field_g} \\
\Box_{\ast} \vp &= -4\pi\alpha(\vp) T_{\ast},
\label{field_phi}
\end{eqnarray}
where $R_{\ast\mu\nu}$ is the Ricci tensor, $\alpha(\vp) \defeq d{\log
  A(\vp) }/d\vp$ (in the language of
\cite{Damour:1993hw,Damour:1996ke}) is the ``scalar-matter coupling
function'' and $\Box_{\ast}$ is the d'Alembertian operator associated
to the metric $g_{\ast\mu\nu}$. GR is obtained in the limit where the
scalar field decouples from matter, i.e.
$\alpha(\vp) \to 0$.
Under the particular choice of the conformal factor (\ref{cf}), this
is equivalent to letting $\beta = 0$. In this paper, all equations
will be derived within the context of scalar-tensor gravity.

Finally, $T_{\ast}^{\mu\nu}$ is the energy-momentum
tensor of the matter fields, defined as
\be
T_{\ast}^{\mu\nu} \defeq  \frac{2}{\sqrt{-g_{\ast}}} \frac{\delta S_{{M}}\left[ \psi_{{M}},A^2(\varphi)g_{\ast\mu\nu} \right]}{\delta g_{\ast\mu\nu}},
\label{t_e}
\ee
and $T_{\ast} \defeq T_{\ast}^{\mu\nu} g_{\ast\mu\nu}$ is its
trace. The energy-momentum tensor in the Jordan frame
$\widetilde{T}^{\mu\nu}$, with trace $\widetilde{T} \defeq
\widetilde{T}^{\mu\nu} \til{g}_{\mu\nu}$, is defined in an analogous
fashion:
\be
\widetilde{T}^{\mu\nu} \defeq  \frac{2}{\sqrt{-\til{g}}} \frac{\delta S_{{M}}\left[ \psi_{{M}},\til{g}_{\mu\nu} \right]}{\delta \til{g}_{\mu\nu}}.
\label{t_j}
\ee
The two energy-momentum tensors (and their traces) are related as
follows:
\begin{eqnarray}
T_{\ast}^{\mu\nu} = A^6(\vp) \widetilde{T}^{\mu\nu}, \quad T_{\ast\mu\nu} = A^2(\vp) \widetilde{T}_{\mu\nu}, \quad
T_{\ast} = A^4(\vp) \widetilde{T}.
\label{t_rel}
\end{eqnarray}
The covariant divergence of the energy-momentum tensor satisfies
\begin{eqnarray}
\nabla_{\ast_{\mu}} T_{\ast}^{\mu\nu} &= \alpha(\vp) T_{\ast} \nabla_{\ast}^{\nu}\vp,
\label{div_t_e} \\
\widetilde{\nabla}_{\mu} \widetilde{T}^{\mu\nu} &= 0,
\label{div_t_j}
\end{eqnarray}
in the Einstein and Jordan frames, respectively.

\subsection{Anisotropic fluids}

An anisotropic fluid with radial pressure $\pJ$, tangential pressure
$\qJ$ and total energy density $\rhoJ$ can be modeled by the
Jordan-frame energy-momentum tensor
\cite{BowersLiang:1974,Doneva:2012rd}
\be
\til{T}_{\mu\nu} = \rhoJ\, \til{u}_{\mu} \til{u}_{\nu} + \pJ \, \til{k}_{\mu}\til{k}_{\nu} + \til{q} \, \til{\Pi}_{\mu\nu},
\label{t_aniso}
\ee
where $\til{u}_{\mu}$ is the fluid four-velocity, $\tilde{k}_{\mu}$ is
a unit radial vector ($\til{k}_{\mu}\til{k}^{\mu} = 1$) satisfying
$\til{u}^{\mu} \til{k}_{\mu} = 0$, and $\til{\Pi}_{\mu\nu} \defeq
\til{g}_{\mu\nu} + \til{u}_{\mu} \til{u}_{\nu} -
\til{k}_{\mu}\til{k}_{\nu}$. $\til{\Pi}_{\mu\nu}$ is a projection
operator onto a two-surface orthogonal to both $\til{u}_{\mu}$ and
$\til{k}_{\mu}$: indeed, defining a projected vector $\tilde{A}^{\mu}
\defeq \til{\Pi}^{\mu\nu} \til{V}_{\nu}$, one can easily verify that
$\til{u}_{\mu} \til{A}^{\mu} = \til{k}_{\mu} \til{A}^{\mu} = 0$.  At
the center of symmetry of the fluid distribution the tangential
pressure $\til{q}$ must vanish, since $\til{k}^{\mu}$ is not defined
there \cite{Doneva:2012rd}. The trace of the Einstein-frame
stress-energy tensor for an anisotropic fluid is
\be\label{Ttrace}
T_{\ast} = A^4(\vp) \left[-(\tilde{\epsilon} - 3\tilde{p})
- 2\left( \pJ - \qJ \right)\right]\,.
\ee
As emphasized by Bowers and Liang \cite{BowersLiang:1974}, $\pJ$ and
$\qJ$ contain contributions from fluid pressures and other possible
stresses inside the star, therefore they should not be confused with
purely hydrostatic pressure.  Additional stresses could be caused, for
instance, by the presence of a solid core
\cite{kippenhahn1990stellar}, strong magnetic fields
\cite{Yazadjiev:2012} or a multi-fluid mixture \cite{Letelier:1980}.
The derivation of a microphysical model for anisotropy is a delicate
issue, so we will adopt a phenomenological approach. We will assume
that $\pJ$ is described by a barotropic EoS, i.e. $\pJ=\pJ(\rhoJ)$.
For brevity in this paper we focus on the APR EoS \cite{Akmal:1998cf},
but we have verified that our qualitative results do not depend on
this choice. The APR EoS supports NS models with a maximum mass $M$
larger than $2.0 \, M_{\odot}$, and therefore it is compatible with
the recent observations of the $M = 1.97 \pm 0.04\, M_{\odot}$ pulsar
PSR J1614-2230~\cite{2010Natur.467.1081D} and of the $M = 2.01\pm
0.04\, M_{\odot}$ pulsar PSR J0348+0432~\cite{Antoniadis:2013pzd}.

The functional form of the anisotropy $\sigmaJ \equiv \pJ -
\qJ$ \cite{BowersLiang:1974,Doneva:2012rd,Glampedakis:2013jya}
depends on microscopic relationships between $\pJ$, $\qJ$ and $\rhoJ$,
that unfortunately are not known. However we can introduce physically
motivated functional relations for $\sigmaJ$ that allow for a smooth
transition between the isotropic and anisotropic regimes. Many such
functional forms have been studied in the literature. As an
application of our general formalism we will consider two of these
phenomenological relations, described below.

\subsubsection{Quasi-local equation of state}

Horvat et al. \cite{Horvat:2010xf} proposed the following quasi-local
equation for $\sigmaJ$:
\be
\sigmaJ \defeq \lambda_{\rm H}\pJ \til{\gamma},
\label{horvat}
\ee
where $\til{\gamma} \defeq 2 \mu(r) / r$. The ``mass function''
$\mu(r)$, defined in Eq.~\eqref{lambda} below, is essentially the mass
contained within the radius $r$, so the quantity $\til{\gamma}$ is a
local measure of compactness, whereas $\lambda_{\rm H}$ is a free
(constant) parameter that controls the degree of anisotropy.

The calculations of \cite{Sawyer:1972cq} show that, if anisotropy
occurs due to pion condensation, $0 \leq \sigmaJ / \pJ \leq 1$,
therefore $\lambda_{\rm H}$ could be of order unity
\cite{Doneva:2012rd}. More recently, Nelmes and Piette
\cite{Nelmes:2012uf} considered NS structure within a model consisting
of a Skyrme crystal, which allows for the presence of anisotropic
strains. They found that $\lambda_{\rm H}$, as defined in
Eq.~\eqref{horvat}, has a nearly constant value $\lambda_{\rm H}
\approx -2$ throughout the NS interior. The nonradial oscillations of
anisotropic stars were studied in \cite{Doneva:2012rd} using the
model of Eq.~$(\ref{horvat})$. Following Doneva and Yazadjiev
\cite{Doneva:2012rd}, we will consider values of $\lambda_{\rm H}$ in
the range $-2 \leq \lambda_{\rm H} \leq 2$.

\subsubsection{Bowers-Liang model}

As a second possibility we will consider the functional form for
$\sigmaJ$ proposed by Bowers and Liang \cite{BowersLiang:1974}, who
suggested the relation\footnote{The factor of 1/3 in
  Eq.~(\ref{BLmod}) is chosen for convenience.
  Also, there is a sign difference between our
  definition of $\sigmaJ$ and the one in
  \cite{BowersLiang:1974}.  Our parameter
  $\lambda_{\rm BL}$ is related with the Bowers-Liang (physically
  equivalent) parameter $C$ by $\lambda_{\rm BL} = - 3 C$.}
\be
\sigmaJ \defeq  \frac{1}{3} \lambda_{\rm BL} \,(\rhoJ + 3\pJ)\, (\rhoJ + \pJ)
\left( 1 - \frac{2\mu}{r} \right)^{-1}\, r^2.
\label{BLmod}
\ee
The model is
based on the following assumptions: (i) the anisotropy should vanish
quadratically at the origin (the necessity for this requirement will
become clear in Sec.~\ref{sec:HT}), (ii) the anisotropy should depend
nonlinearly on $\pJ$, and (iii) the anisotropy is (in part)
gravitationally induced. The parameter $\lambda_{\rm BL}$ controls the
amount of anisotropy in the fluid.

This ansatz was used in \cite{BowersLiang:1974} to obtain an exact
solution for incompressible stars with $\rhoJ=\rhoJ_0=$ constant. In
their simple model, the requirement that equilibrium configurations
should have finite central pressure $\pJ_{\tc}$ implies that
$\lambda_{\rm BL} \geq - 2$.
The Newtonian limit of the Bowers-Liang ansatz was also considered in
a recent study of the correspondence between superradiance and tidal
friction \cite{Glampedakis:2013jya}.
In our calculations we will assume that $-2 \leq \lambda_{\rm BL} \leq
2$.

\section{Stellar structure in the slow-rotation approximation}
\label{sec:HT}

In this Section we approximate the metric of a slowly, rigidly
rotating, anisotropic star following the seminal work by Hartle and
Thorne \cite{Hartle:1967he,Hartle:1968si}. The idea is to consider the
effects of rotation as perturbations of the spherically symmetric
background spacetime of a static star. We generalize the results of
\cite{Hartle:1967he,Hartle:1968si} (in GR) and \cite{Damour:1996ke}
(in scalar-tensor theory) to account for anisotropic fluids up to
first order in rotation, so we can
study how anisotropy and scalarization affect the moment of inertia of
the star and the dragging of inertial frames.

We remark that the moment of inertia $I$, the star's uniform angular
velocity $\Omega$ and the angular momentum $J \defeq I \Omega$ are the
same in the Jordan and Einstein frames
(cf. \cite{Damour:1996ke,Pani:2014jra}). Therefore, to simplify the
notation, we will drop asterisks and tildes on these quantities.
Working at order ${\cal O}(\Omega)$, the line element of a
stationary axisymmetric spacetime
in the Jordan frame reads
\begin{eqnarray}
d{\til{s}^2} = A^2(\vp) \Big[&- e^{2\Phi(r)}dt^2 + e^{2 \Lambda(r)}dr^2 + r^2 d\theta^2  \nonumber \\
& +\, r^2\sin^2{\theta}\,d\phi^2 - 2  \,{\omega}(r,\theta) r^2 \sin^2\theta\, dt \, d\phi \Big],
\label{metric_rot_j}
\end{eqnarray}
where
\be
e^{-2\Lambda(r)} \defeq 1 - \frac{2\mu(r)}{r},
\label{lambda}
\ee
$\mu(r)$ is the mass function and
${\omega}(r,\theta)\sim {\cal O}
(\Omega)$ is the angular velocity acquired by a particle falling from
infinity as measured by a static asymptotic observer
\cite{Hartle:1967he}.

The four-velocity of the rotating fluid is such that $\uJ_{\mu}
\uJ^{\mu} = -1$, and it has components \cite{Hartle:1967he}
\begin{eqnarray}
\uJ^0 &= \left[ - (\gJ_{00} + 2\OmegaJ\gJ_{03} + \OmegaJ^2 \gJ_{33}) \right]^{-1/2}, \\
\uJ^1 &= \uJ^2 = 0, \\
\uJ^3 &=  \OmegaJ \uJ^0.
\label{four_vec_rot}
\end{eqnarray}
Using (\ref{metric_rot_j}), at first order in the slow-rotation
parameter we obtain:
\be
\uJ^{\mu} = A^{-1}(\vp)\left( e^{-\Phi}, 0, 0,\Omega \, e^{-\Phi}\right).
\ee

Following the standard procedure
\cite{Misner1973,Hartle:1967he,2014JCAP...10..006S}, the field
equations (\ref{field_g}), (\ref{field_phi}) and (\ref{div_t_e}) with
the metric given by \eqref{action} yield the following set of ordinary
differential equations:
\begin{eqnarray}
\frac{d\mu}{dr} &= 4 \pi A^4(\vp) r^2 \rhoJ  + \frac{1}{2}r(r-2\mu) \psi^2, \label{dmu} \\
\frac{d\Phi}{dr} &= 4 \pi A^4(\vp) \frac{r^2 \pJ}{r-2\mu} + \frac{1}{2}r\psi^2 + \frac{\mu}{r(r-2\mu)}, \label{dphi} \\
\frac{d\psi}{dr} &= 4 \pi A^4(\vp) \frac{r}{r-2\mu}\left[ \alpha(\vp)(\rhoJ - 3\pJ) + r(\rhoJ - \pJ)\psi \right]  \nonumber \\
& - \frac{2(r-\mu)}{r(r-2\mu)}\psi + 8 \pi  A^4(\vp) \alpha(\vp)\frac{r \til{\sigma}}{r-2\mu},\label{dpsi} \\
\frac{d\pJ}{dr} &= - (\rhoJ + \pJ) \left[ \frac{d\Phi}{dr} + \alpha(\vp)\psi \right] - 2 \til{\sigma} \left[ \frac{1}{r} + \alpha(\vp)\psi \right], \label{dp}\\
\frac{d\varpi}{dr} &= 4 \pi  A^4(\vp) \frac{r^2}{r-2\mu} (\rhoJ + \pJ)\left( \varpi + \frac{4\omegab}{r} \right)
+  \left( r \psi^2 - \frac{4}{r} \right) \varpi \nonumber \\
&+ 16 \pi A^4(\vp) \frac{r\tilde{\sigma}}{r-2\mu} \omegab,
\label{domegab}
\end{eqnarray}
where we defined $\psi \defeq d\vp/dr$, $\varpi \defeq d\omegab /dr$,
and $\omegab\defeq \Omega-\omega$. The equations above reduce to the
Tolman-Oppenheimer-Volkoff (TOV) equations for anisotropic stars in GR
\cite{BowersLiang:1974} when $\alpha \to 0$, to the results of
\cite{Damour:1993hw} in the isotropic limit $\sigmaJ \to 0$, and to
the usual TOV equations when both quantities are equal to zero
\cite{Misner1973}. In the GR limit, our frame-dragging equation
(\ref{domegab}) agrees with Bayin's \cite{Bayin:1982vw}
result\footnote{In principle, as mentioned in the introduction,
  rotation may induce anisotropy. Therefore the Horvat et al. and
  Bowers-Liang models for $\sigmaJ$ should contain terms proportional
  to $\Omega$. However, Eq.~(\ref{domegab}) implies that such terms in
  $\sigmaJ$ would lead to corrections of second order in the angular
  velocity $\Omega$. These corrections are beyond the scope of the
  ${\cal O}(\Omega)$ approximation considered in our work.}.

To obtain the interior solution we integrate the generalized TOV
equations (\ref{dmu})-(\ref{domegab}) from a point $r_{\tc}$ close to
the stellar center $r = 0$ outwards up to a point $r = r_{\ts}$ where
the pressure vanishes, i.e. $\pJ(r_{\ts}) = 0$.
This point specifies the Einstein-frame radius $R_{\ast} \defeq
r_{\ts}$ of the star. If $\varphi_{\ts}=\varphi(r_{\ts})$, the
Jordan-frame radius $\tilde{R}$ is
\be
\tilde{R} = A(\varphi_{{\ts}})\, R_{\ast}.
\ee

In practice, to improve numerical stability, given $\rhoJ_{\tc}$,
$\Phi_{\tc}$, $\varphi_{\tc}$ and $\mu_{\tc}$ (where the subscript c
means that all quantities are evaluated at $r = 0$) we use the
following series expansions:
\begin{eqnarray}
\mu &= \frac{4}{3} \pi A_{\rm c}^4 \rhoJ_{\rm c}  r^3 + {\cal O}(r^4),\nn\\
\Phi &= \Phi_{\tc} + \frac{2}{3} \pi A_{\rm c}^4 \left(\rhoJ_{\rm c}+3 \pJ_{\rm c}\right)  r^2 + {\cal O}(r^4),\nn\\
\pJ &= \pJ_{\tc}+\frac{2}{3} \pi  r^2 A_{\rm c}^4 \left(\rhoJ_{\rm c}+\pJ_{\rm c}\right) \left[3 \pJ_{\rm c} \left(\alpha_{\rm c}^2-1\right)-\rhoJ_{\rm c}\left(\alpha_{\rm c}^2+1\right)\right]+\nn\\
&- \frac{1}{3} r^2 (2 r \sigma_3+3 \sigma _2) + {\cal O}(r^4), \nn\\
\vp &= \vp_{\rm c} + \frac{2\pi}{3} A^4_{\tc} \alpha_{\tc} (\rhoJ_{\tc} - 3\pJ_{\tc})r^2 + {\cal O}(r^4), \nn\\
\omegab &= \omegab_{\tc} + \frac{8\pi}{5} A^4_{\tc} \omegab_{\tc} (\rhoJ_{\tc} + \pJ_{\tc}) r^2 + {\cal O}(r^4), \nn\\
\til{\sigma } &=\sigma_2 r^2+\sigma_3 r^3+{\cal O}(r^{4}),
\end{eqnarray}
where $\sigma_2$ and $\sigma_3$ depend on the particular anisotropy
model.

In the vacuum exterior we have $\pJ = \rhoJ = \sigmaJ =
0$. Eqs.~\eqref{dmu}--\eqref{dpsi} must be integrated outwards
starting from the stellar radius to obtain the stellar mass, angular
momentum and scalar charge.
For large $r$ we can expand the relevant functions as follows:
\begin{eqnarray}
\mu(r) &= M - \frac{Q^2}{2r} - \frac{M Q^2}{2r^2} + {\cal O}(r^{-3}) \label{muinf} \\
e^{2\Phi} &= 1-\frac{2M}{r}+{\cal O}(r^{-3}), \label{phiinf}\\
\varphi(r) &= \varphi_\infty + \frac{Q}{r} + \frac{MQ}{r^2} + {\cal O}(r^{-3}), \label{vpinf}\\
\omegab(r) &=\Omega- \frac{2 J}{r^3}+{\cal O}(r^{-4}), \label{winf}
\end{eqnarray}
where $M$ is the Arnowitt-Deser-Misner (ADM) mass of the NS, $Q$ is
the scalar charge, $J$ is the star's angular momentum and
$\varphi_\infty$ is the (constant) cosmological value of the scalar
field, here assumed to be zero. Under this assumption the mass $M$ is
the same in the Jordan and Einstein frames \cite{Pani:2014jra}.  By
matching the numerical solution integrated from the surface of the
star with the asymptotic expansions \eqref{muinf}--\eqref{winf} we can
compute $M$, $Q$ and $J$.

We compute the moment of inertia of the star $I$ in two equivalent
ways. The first method consists of extracting the angular momentum as
described above and using
\be
I=\frac{J}{\Omega}.
\label{eq:ine1}
\ee
In alternative, we can compute $I$ through an integral within the
star. Combining Eqs.~(\ref{lambda}), (\ref{dmu})-(\ref{dphi}) and
(\ref{domegab}) we obtain the following integral expression:
\begin{eqnarray}
I = \frac{8 \pi }{3} \int_0^{R_{\ast}} A^4(\vp) e^{\Lambda-\Phi } r^4 (\rhoJ + \pJ) \left( 1 - \frac{\sigmaJ}{\rhoJ + \pJ} \right) \left( \frac{\omegab}{\Omega} \right) \, dr
\label{inertia}
\end{eqnarray}
(see Appendix \ref{sec:I} for details). As $A(\vp) \to 1$ and $\sigmaJ
\to 0$ we recover Hartle's result \cite{Hartle:1967he}, and in the
isotropic limit $\sigmaJ \to 0$ we match the result of
\cite{2014JCAP...10..006S}. The numerical values of $I$ obtained with
\eqref{eq:ine1} and \eqref{inertia} are in excellent agreement.

For each stellar model we also calculate the {\it baryonic mass}
$\Mb$, defined as \cite{Damour:1993hw}
\be
\Mb \defeq 4 \pi \tilde{m}_{\rm{b}} \int_0^{R_{\ast}}\tilde{n}\, A^3(\vp) \frac{r^2}{\sqrt{1 - 2\mu / r}} \, dr,
\ee
where $\tilde{m}_{\rm{b}} = 1.66 \times 10^{-24}$ g is the atomic mass
unit and $\tilde{n}$ is the baryonic number density.

\begin{figure}[htb]
\includegraphics[width=\columnwidth]{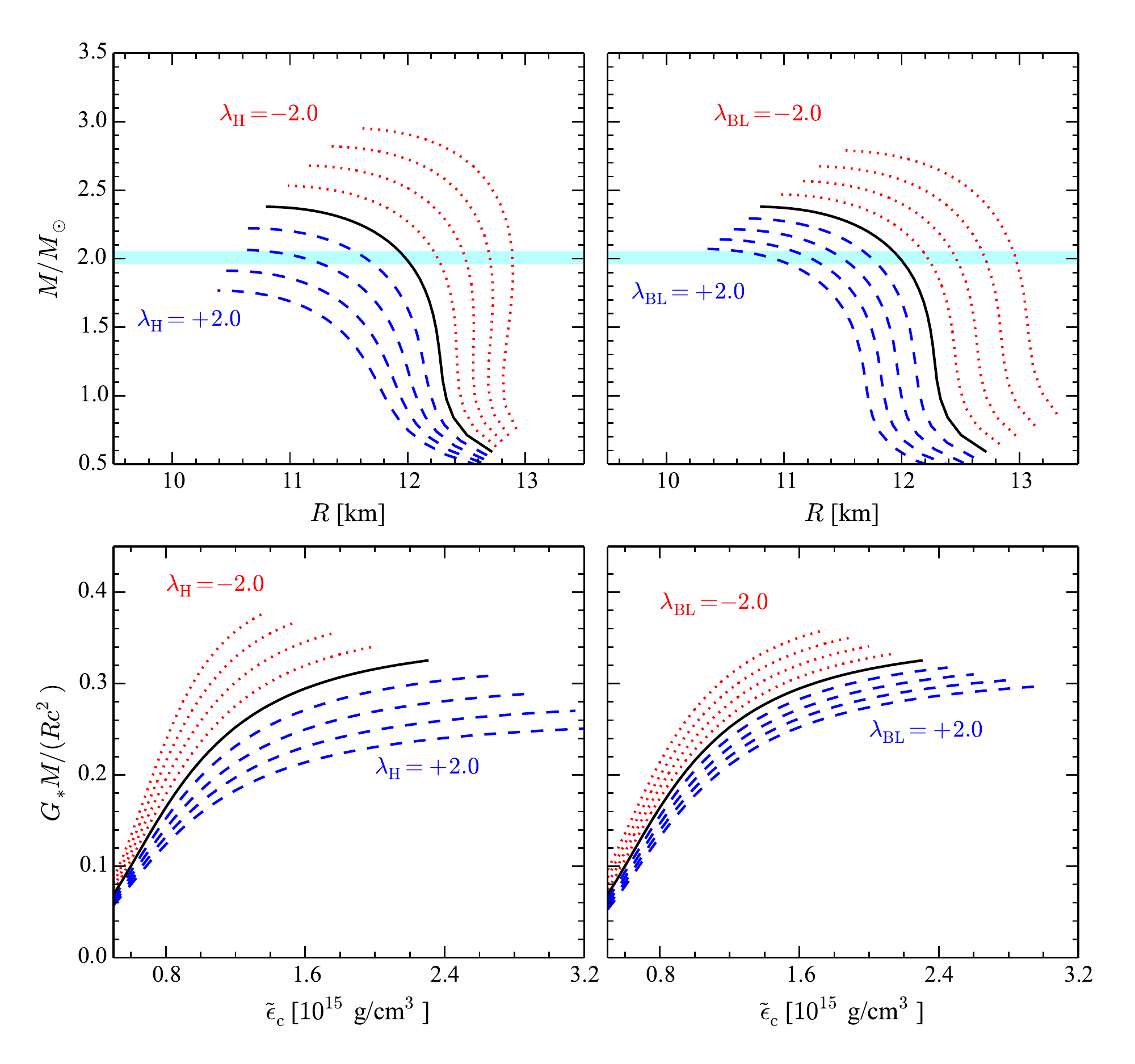}
\caption{Mass-radius relation (top panels) and dimensionless
  compactness $G_{\ast} M/Rc^2$ as a function of the central density
  (bottom panels) for anisotropic stars in GR using EoS APR. In the
  left panels we use the quasi-local model of \cite{Horvat:2010xf}; in
  the right panels, the Bowers-Liang model
  \cite{BowersLiang:1974}. Different curves correspond to increasing
  $\lambda_{\rm H}$ (or $\lambda_{\rm BL}$) in increments of $0.5$
  between $-2$ (top) and $2$ (bottom). The shaded blue bar corresponds
  to the mass $M = 2.01\pm 0.04\, M_{\odot}$ of PSR J0348+0432
  \cite{Antoniadis:2013pzd}.}
\label{mass_radius_gr}
\end{figure}

\section{Numerical results}
\label{sec:results}

The tools developed so far allow us to investigate the effect of
anisotropy on the bulk properties of rotating stars. In Section
\ref{aniso_gr} we will focus on slowly rotating stars in GR. To the
best of our knowledge -- and to our surprise -- rotating anisotropic
stars have not been studied in the GR literature, with the only
exception of a rather mathematical paper by Bayin
\cite{Bayin:1982vw}. In Section \ref{aniso_st} we extend our study to
scalar-tensor theories. Our main motivation here is to understand
whether anisotropy may increase the critical value $\beta=\beta_{\rm
  crit}$ above which spontaneous scalarization cannot happen,
and therefore allow for observationally interesting modifications to
the structure of NSs that would still be compatible with the stringent
bounds from binary pulsars \cite{Freire:2012mg,Wex:2014nva}.

\subsection{The effect of anisotropy in GR} \label{aniso_gr}

In the top panels of Figure~\ref{mass_radius_gr} we show the mass-radius
relation for
anisotropic NS models in GR. All curves are truncated at the central
density corresponding to the maximum NS mass, because models with
larger central densities are unstable to radial perturbations
\cite{HTWW1965,Misner1973}. Solid lines correspond to $\sigmaJ = 0$,
i.e. the isotropic fluid limit. The horizontal shaded band in the
upper panels represents the largest measured NS mass
$M = 2.01\pm 0.04\, M_{\odot}$ (PSR J0348+0432: cf.~\cite{Antoniadis:2013pzd}).

Recall that $\sigmaJ = \pJ- \qJ$ is proportional to $\lambda_{\rm
  H}$ and $\lambda_{\rm BL}$ (with a positive proportionality
constant) in both models, and that $\pJ$ and $\qJ$ represent the
``radial'' and ``tangential'' pressures, respectively.
Therefore positive values of $\lambda_{\rm H}$ and $\lambda_{\rm BL}$
mean that the radial pressure is larger than the tangential pressure
(dashed lines); the opposite is true when the anisotropy parameters
are negative (dotted lines).

The trend in the top panels of Figure~\ref{mass_radius_gr} is clear:
for both anisotropy
models, positive (negative) anisotropy parameters yield smaller
(larger) radii at fixed mass, and smaller masses at fixed radius. The
lower panels of Figure~\ref{mass_radius_gr}
show that the stellar compactness $G_{\ast}M/(Rc^2)$ decreases (for a
given EoS and fixed central density) as the anisotropy degree
increases.
Nuclear matter EoSs are usually ordered in terms of a ``stiffness''
parameter, with stiffer EoSs corresponding to larger sound speeds
(more incompressible matter) in the stellar interior, and larger
values of the compactness $M/R$. The qualitative effect of increasing
anisotropy (with our sign conventions) is {\em opposite} (for a given
EoS) to the qualitative effect of increasing stiffness.

\begin{figure}[h]
\includegraphics[width=\columnwidth]{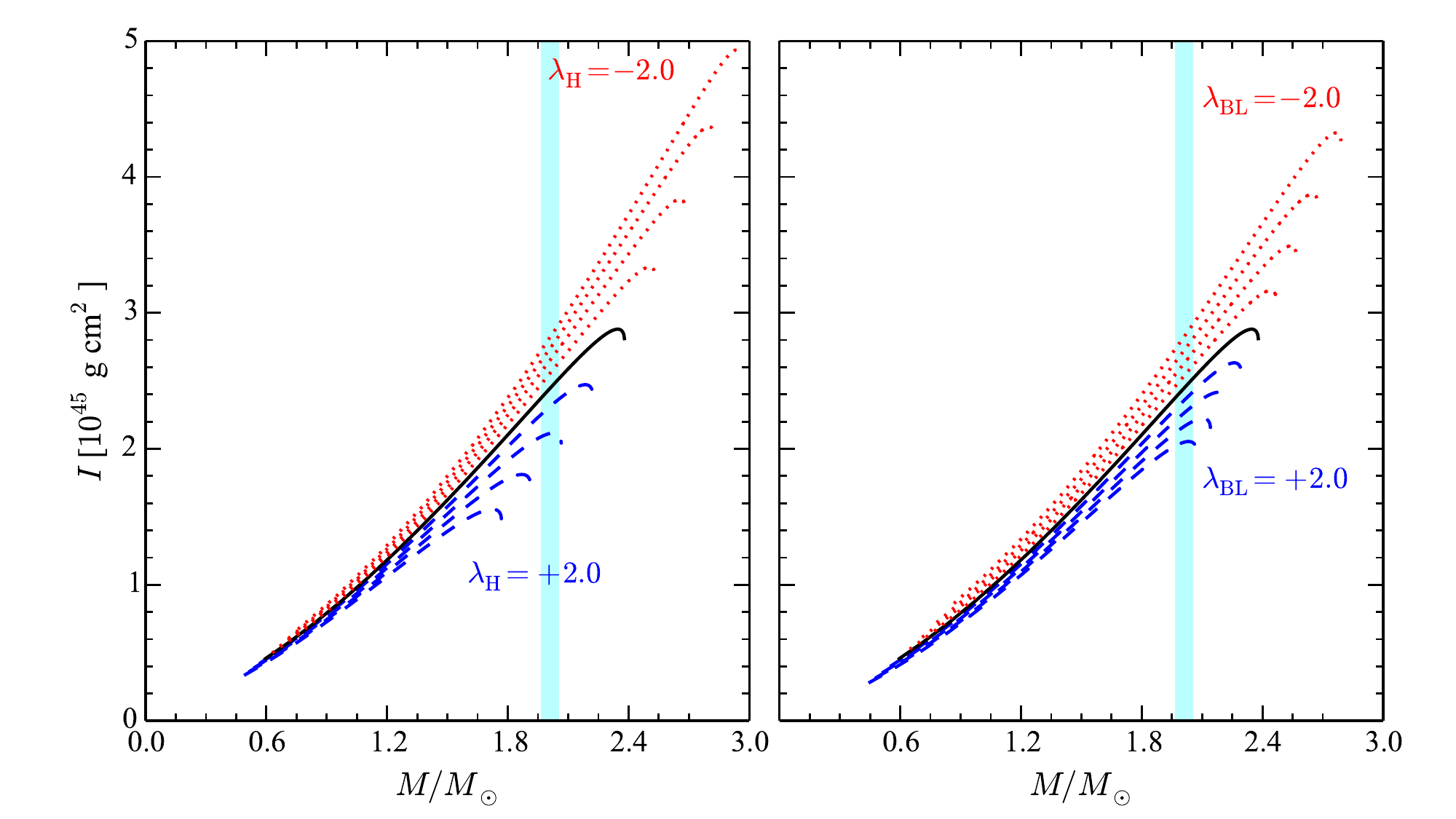}
\caption{The moment of inertia $I$ as function of the mass $M$ for
  anisotropic stars in GR using EoS APR, increasing $\lambda_{\rm H}$
  (or $\lambda_{\rm BL}$) in increments of $0.5$ between $-2$ (top
  curves) and $2$ (bottom curves). As in Figure \ref{mass_radius_gr},
  the vertical shaded region marks the largest measured NS mass
  \cite{Antoniadis:2013pzd}.  }
\label{i_mass_gr}
\end{figure}

Figure~\ref{i_mass_gr} is, to our knowledge, the first calculation of
the effect of anisotropy on the moment of inertia $I$.
As in Figure~\ref{mass_radius_gr}, solid lines corresponds to the
isotropic limit. In the right panel we use the quasi-local model of
\cite{Horvat:2010xf}; in the left panel, the Bowers-Liang model
\cite{BowersLiang:1974}. Hypothetical future
observations of the moment of inertia of star A, from the double
pulsar PSR J0737-3039
\cite{Lyne:2004cj,Lattimer:2004nj,Kramer:2009zza}, or preferably from
large-mass NSs, may be used to constrain the degree of anisotropy
under the assumptions that GR is valid and that the nuclear EoS is
known.
\begin{figure}[h]
\includegraphics[width=\columnwidth]{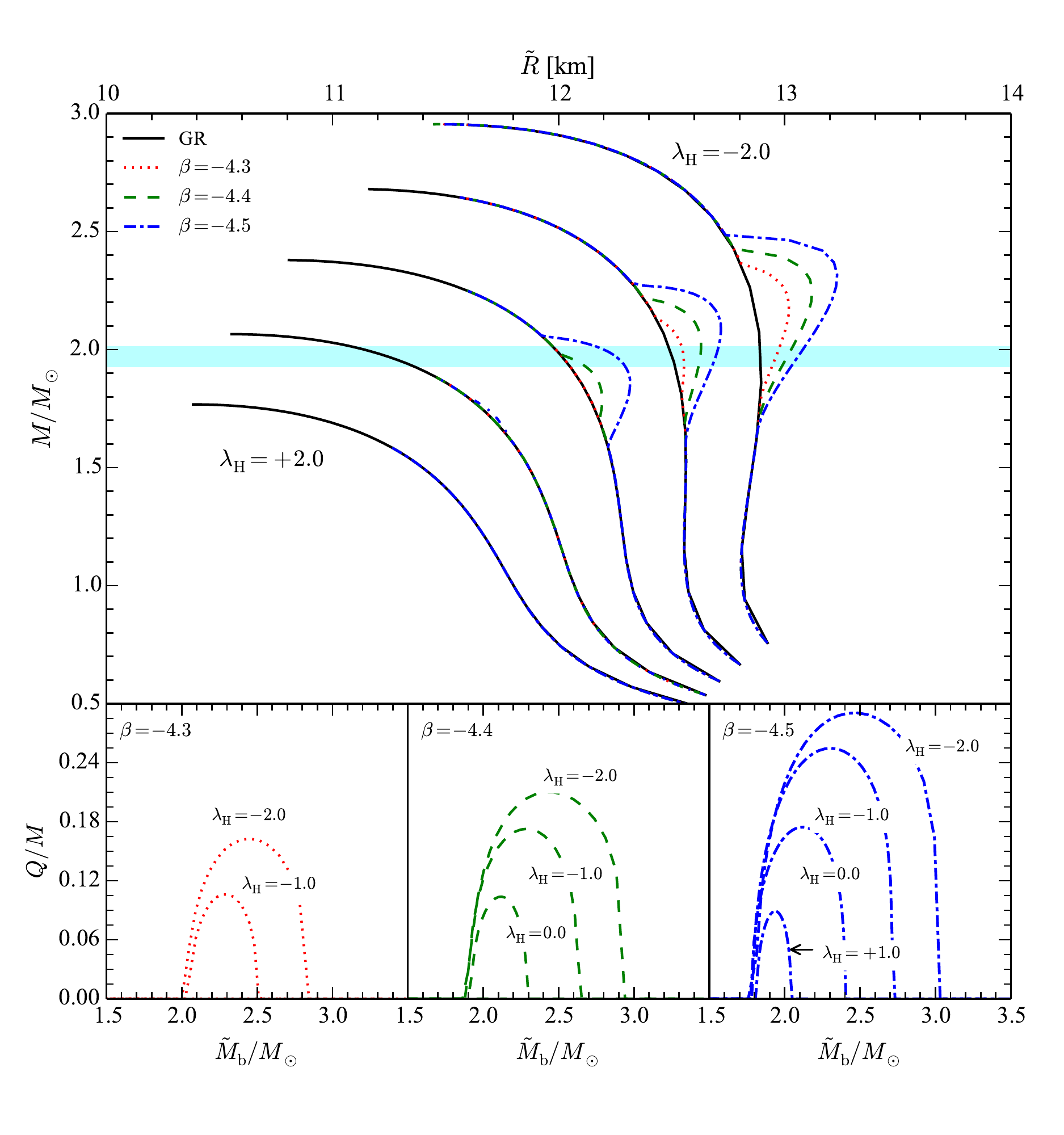}
\caption{Spontaneous scalarization in the quasi-local model of Horvat
  et al. \cite{Horvat:2010xf}. See the main text for details.}
\label{fig:ro_scalar_QL}
\end{figure}

\begin{figure}[h]
\includegraphics[width=\columnwidth]{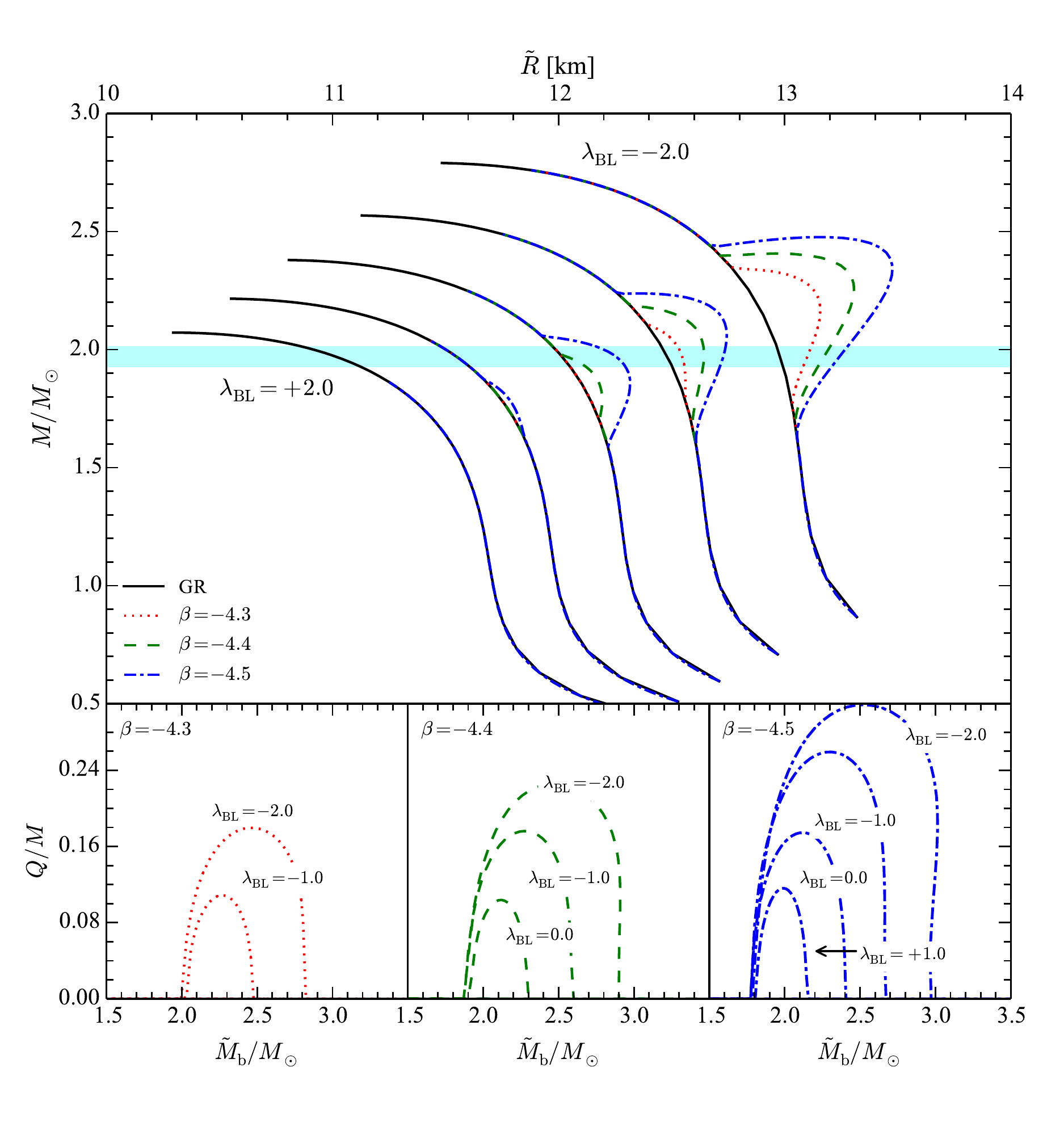}
\caption{Same as Figure~\ref{fig:ro_scalar_QL}, but for the
  Bowers-Liang anisotropy model \cite{BowersLiang:1974}.}
\label{fig:ro_scalar_BL}
\end{figure}

\subsection{The effect of anisotropy on spontaneous scalarization}
\label{aniso_st}

In Figures~\ref{fig:ro_scalar_QL} and \ref{fig:ro_scalar_BL} we
display the properties of nonrotating, spontaneously scalarized stars
within the anisotropy models of Horvat et al. \cite{Horvat:2010xf} and
Bowers-Liang \cite{BowersLiang:1974}, respectively.
The main panel in each Figure shows the mass-radius relation as the
anisotropy parameter increases (in increments of 1, and from top to
bottom) in the range $-2\leq \lambda_{\rm H}\leq 2$
(Figure~\ref{fig:ro_scalar_QL}) or $-2\leq \lambda_{\rm BL}\leq 2$
(Figure~\ref{fig:ro_scalar_BL}). Solid lines correspond to the GR
limit; dotted, dashed and dash-dotted lines correspond to
$\beta=-4.3,\,-4.4$ and $-4.5$, as indicated in the legend. The lower panels
show the scalar charge $Q/M$ as a function of the baryonic mass. In
each of these panels we plot the scalar charge for a fixed value of $\beta$ and
different anisotropy parameters.

For isotropic EoSs in GR, Harada \cite{Harada:1998ge} used catastrophe
theory to show that scalarization is only possible when $\beta\lesssim
-4.35$.
We find that scalarization can occur for larger values of $\beta$ in
the presence of anisotropy. For example, for a value of $\lambda_{\rm
  H}\sim -2$ (compatible with the Skyrme model predictions of
\cite{Nelmes:2012uf}) scalarization is possible when $\beta\simeq
-4.15$, and for $\beta\simeq -4.3$ scalarization produces rather large
($\approx 10\%$) deviations in the mass-radius relation. This
qualitative conclusion applies to both anisotropy models considered
by us. The lower panels
show that: (i) for fixed $\beta$ (i.e., for a fixed theory) and for a
fixed central density, the ``strength'' of scalarization -- as
measured by the scalar charge of the star -- increases for large
negative $\lambda$'s, i.e. when the tangential pressure is
significantly larger than the radial pressure, for both anisotropy
models; (ii) scalarization occurs in a much wider range of baryonic
masses, all of which are compatible with the range where anisotropy
would be expected according to the Skyrme model predictions of
\cite{Nelmes:2012uf}. These calculations are of course preliminary and
should be refined using microphysical EoS models.
However, let us remark once again that the scalarization threshold in
the absence of anisotropy is to a very good approximation
EoS-independent, and stars only acquire significant scalar charge
when $\beta<-4.35$ (as shown in~\cite{Harada:1998ge} and in
Figure~\ref{cri_bet_eos} below).
In the admittedly unlikely event that binary pulsar observations were
to hint at scalarization with $\beta>-4.35$, this would be strong
evidence for the presence of anisotropy\footnote{An important caveat
  here is that {\em fast} rotation can also strengthen the effects of
  scalarization: according to \cite{Doneva:2013qva}, scalarization can
  occur for $\beta<-3.9$ for NSs spinning at the mass-shedding
  limit. However the NSs found in binary pulsar systems are relatively
  old, as they are expected to be spinning well below the
  mass-shedding limit, where the slow-rotation approximation works
  very well \cite{Berti:2004ny}.} and even lead to experimental
constraints on the Skyrme model and QCD.

\begin{figure}[h]
\includegraphics[width=\columnwidth]{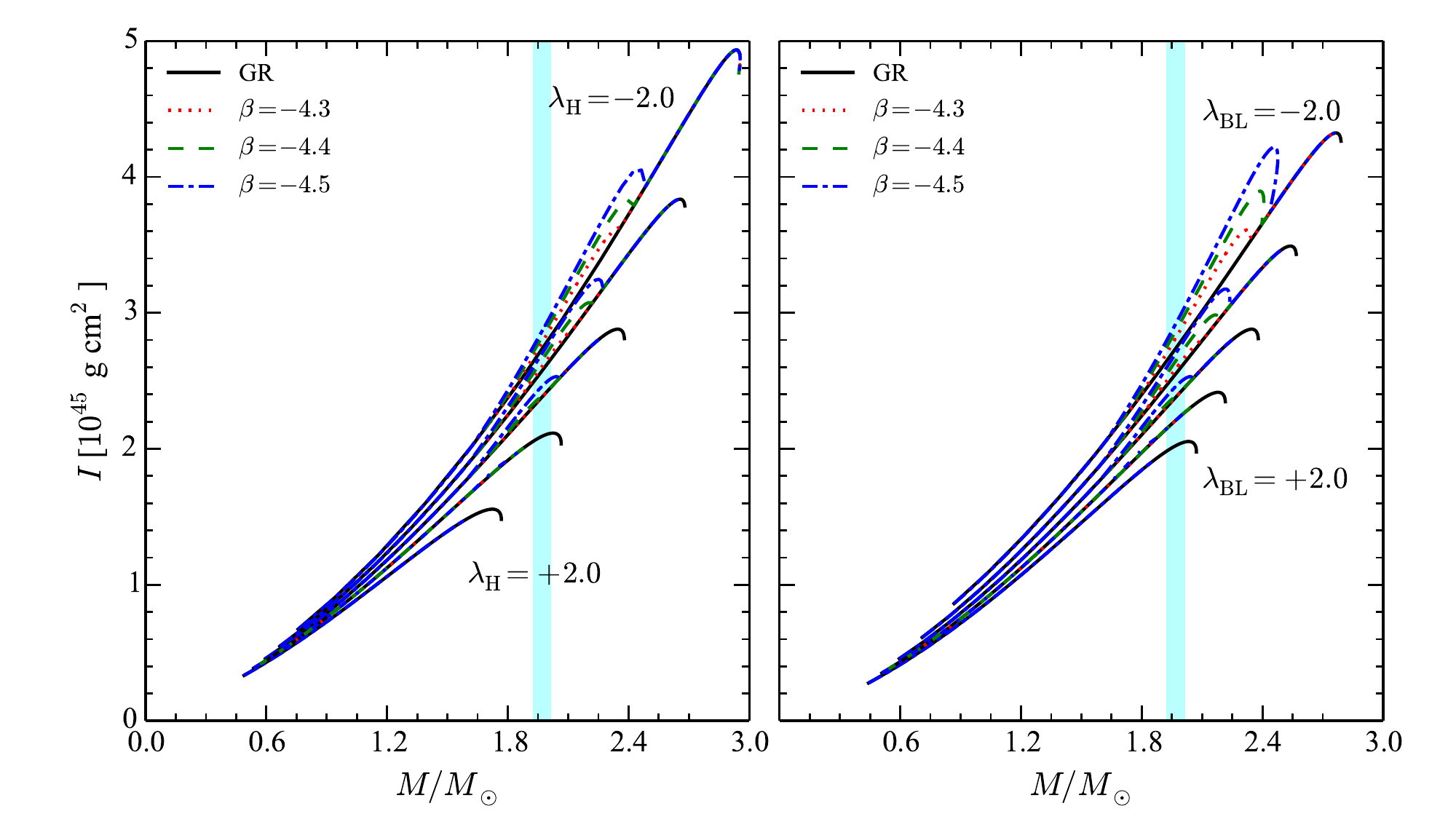}
\caption{Same as Figure~\ref{i_mass_gr}, but in scalar-tensor theories
  with different values of $\beta$.}
\label{i_mass_st}
\end{figure}

As in Figure~\ref{i_mass_gr}, in the left panel of
Figure~\ref{i_mass_st} we show the moment of inertia as a function of
the stellar mass for the quasi-local model of \cite{Horvat:2010xf},
while the right panel refers to the Bowers-Liang model
\cite{BowersLiang:1974}.
Solid lines corresponds to the GR limit for different anisotropy
parameters. Unsurprisingly, the largest modifications to the moment of
inertia occur for large negative $\lambda$'s, and they follow the same
trends highlighted in our discussion of the mass-radius relation.

\subsection{Critical scalarization point in the linearized approximation}
\label{sec:critical_mass}

The condition for spontaneous scalarization to occur can be found in a
linearized approximation to the scalar-field equation of motion. The
idea is that {\em at the onset} of scalarization the scalar field must
be small, so we can neglect its backreaction on the geometry and look
for bound states of the scalar field by dropping terms quadratic in
the field \cite{Damour:1993hw,Damour:1996ke}. Here we study general
conditions for the existence of bound states in the linearized regime,
and we show that (as expected based on the previous argument) the
linearized theory does indeed give results in excellent agreement with
the full, nonlinear calculation.

Redefining the scalar field as
$\varphi(t,r)=r^{-1}\Psi(r) e^{-i \nu t}$ and neglecting terms
${\cal O}(\varphi^2)$, Eq.~(\ref{field_phi}) can be written as a
Schr\"odinger-like equation:
\be
\frac{d^2\Psi}{dx^2}+\left[\nu^2-{V}_{\rm eff}(x) \right]\Psi=0,
\label{eq:scalar_mod}
\ee
where the tortoise radial coordinate $x$ is defined by $dx \defeq
dr\,e^{-\Phi}/\sqrt{1-2\mu/r}$. The effective potential is
\be
{V}_{\rm eff}(r)\equiv e^{2\Phi}\left[\mu_{\rm eff}^2(r)+\frac{2\mu}{r^3}+4\pi(\pJ-\rhoJ)\right],
\label{eq:pot_mod}
\ee
where we have introduced an effective (position-dependent) mass
\be
\mu_{\rm eff}^2(r)\defeq -4\pi \beta T_{\ast}.
\ee
Eq.~\eqref{eq:scalar_mod} with the potential \eqref{eq:pot_mod} is a
wave equation for a scalar field with effective mass $\mu_{\rm
  eff}$. From Eq.~\eqref{Ttrace} we see that anisotropy affects the
effective mass (and therefore the scalarization threshold) because
$T_{\ast}$ contains a term proportional to $\tilde{\sigma}$, that in
turn is proportional to either $\lambda_{\rm H}$ or $\lambda_{\rm
  BL}$: cf. Eqs.~\eqref{horvat} and \eqref{BLmod}.
The case of spontaneous scalarization around black holes (studied in
\cite{Cardoso:2013opa,Cardoso:2013fwa}) can be recovered by setting
$\rhoJ=\pJ=0$.

The scalarization threshold can be analyzed by looking for the
zero-energy ($\nu\sim 0$) bound state solutions of
Eq.~\eqref{eq:scalar_mod}. In this case, the scalar field satisfies
the following boundary conditions:
\be
\Psi\sim\left\{
\begin{array}{ll}
    \vp_{\rm c} r & {\rm as}\quad r\to 0,\\
    \vp_{\infty}  & {\rm as}\quad r\to \infty,
\end{array}\right.
\ee
and we impose $\Psi'(r\to\infty)=0$, where the prime denotes derivative
with respect to $r$. To obtain the scalarization threshold
we integrate Eq.~\eqref{eq:scalar_mod} outwards, starting from $r=0$,
with the above boundary conditons. Since the equation is linear,
$\vp_{\rm c}$ is arbitrary. At infinity we impose that the first
derivative of $\Psi$ with respect to $r$ must be zero. This is a
two-point boundary value problem that can be solved with a standard
shooting method to find the critical value of the central density
$\rhoJ_{\rm c}$ for which the above conditions are satisfied, given fixed
values of $\beta$ and $\lambda_{\rm H}$ (or $\lambda_{\rm BL}$).  The
solution is some
\be
\rhoJ_{\rm i}=\rhoJ_{\rm i} (\beta),
\ee
where $\rhoJ_{\rm i}$ is the smallest critical density at which
scalarization can occur for the given $\beta$. The largest critical
density producing scalarization can be similarly obtained by looking
for zero-energy bound state solutions to find some
\be
\rhoJ_{\rm f}=\rhoJ_{\rm f} (\beta).
\ee
It can be shown that in these two regimes (i.e., at the starting and
ending points of the scalarization regime) the derivative of
$\Psi'(r\to \infty)$ with respect to $\rhoJ_{\rm c}$ has opposite signs:
\be
\frac{\pa}{\pa \rhoJ_{\rm c}} \Psi'(r\to \infty)\left\{
\begin{array}{l l}
  <0 & {\rm for} \quad \rhoJ_{\rm c}=\rhoJ_{\rm i},\\
  >0 & {\rm for} \quad \rhoJ_{\rm c}=\rhoJ_{\rm f}.
\end{array}\right.
\ee

\begin{figure}[h]
\includegraphics[width=\columnwidth]{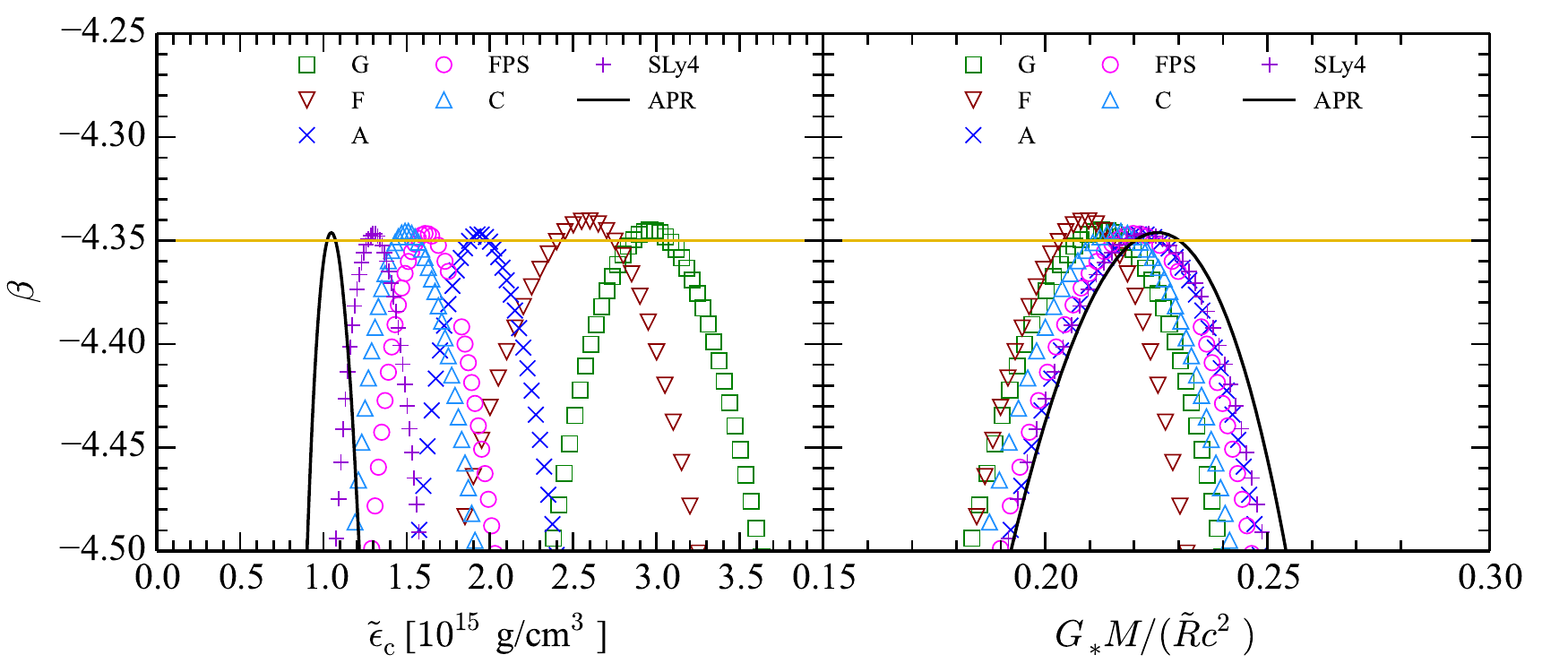}
\caption{Critical $\beta$ for scalarization as a function of the
  central density (left panel) and of the stellar compactness (right
  panel) for nonrotating NS models constructed using different
  nuclear-physics based EoSs, in the absence of anisotropy.}
\label{cri_bet_eos}
\end{figure}

As a warm-up, in Figure~\ref{cri_bet_eos} we compute the scalarization
threshold for nonrotating isotropic stars with several nuclear-physics
based EoSs. The original references for the subset of EoSs used here
can be found in \cite{Kokkotas:2000up} (the one exception is SLy4:
cf.~\cite{Douchin:2001sv}).  The EoSs are sorted by stiffness, with APR EoS
being the stiffest and G EoS the softest in our catalog.
As a trend, for stiffer EoSs scalarization occurs at lower values of
the central densities and at higher values of the compactness.
The most remarkable fact is that the value $\beta=\beta_{\rm max}$
above which scalarization cannot occur is very narrow: it ranges from
$\beta_{\rm max} = -4.3462$ for APR EoS to $\beta_{\rm max} = -4.3405$
for F EoS \cite{Arponen:1972zz}.
This is consistent with Harada's study based on catastrophe
theory, that predicts a threshold value $\beta_{\rm max}\simeq -4.35$
(horizontal line in the figure) in the absence of anisotropy
\cite{Harada:1998ge} (see also \cite{Novak:1998rk}).

\begin{table}
\center
\caption{Critical density values obtained through the linearized
  theory and the full nonlinear equations for APR EoS, different
  values of the Horvat et al. anisotropy parameter $\lambda_{\rm H}$
  and $\beta=-4.5$: for these choices of parameters, the solution is
  scalarized if $\rhoJ_{\rm i}< \rhoJ_{\rm c} <\rhoJ_{\rm f}$. The last column
  lists the critical value $\beta=\beta_{\rm max}$ above which
  scalarization is not possible.}
\begin{tabular}{ c  c  c  c  c c}
\hline
\hline
& Linearized & & Full nonlinear & & \\
$\lambda_{\rm H}$ & $\rhoJ_{\rm i} ({\rm g\, cm}^{-3})$ & $\rhoJ_{\rm f} ({\rm g\, cm}^{-3})$ & $\rhoJ_{\rm i} ({\rm g\, cm}^{-3})$ & $\rhoJ_{\rm f} ({\rm g\, cm}^{-3})$ & $\beta_{\rm max}$\\
\hline
-2 & $6.983\times 10^{14}$ & $9.141\times 10^{14}$ &   $6.980\times 10^{14}$& $9.140\times 10^{14}$ & -4.150\\
-1 & $7.819\times 10^{14}$ & $1.053\times 10^{15}$ & $7.817\times 10^{14}$& $1.053\times 10^{15}$ & -4.239\\
0 & $9.021\times 10^{14}$ & $1.216\times 10^{15}$ &  $9.021\times 10^{14}$ & $1.216\times 10^{15}$ & -4.346\\
1 & $1.127\times 10^{15}$ & $1.340\times 10^{15}$ & $1.126\times 10^{15}$& $1.341\times 10^{15}$ & -4.471\\
\hline\hline
\end{tabular}
\label{tab:scalar}
\end{table}

In Table \ref{tab:scalar} we compare the values for $\rhoJ_{\rm i}$
and $\rhoJ_{\rm f}$ computed using (i) the linearized method described
in this Section, and (ii) the full nonlinear set of equations for
anisotropic models constructed using the APR EoS. The results agree
remarkably well, showing that the onset of scalarization can be
analyzed to an excellent degree of accuracy by neglecting the
backreaction effects of the scalar field on the geometry.
The last column of Table \ref{tab:scalar} lists $\beta_{\rm max}$, the
value of $\beta$ above which scalarization cannot happen. We do not
present results for $\lambda_{\rm H}=2$ because the resulting
$\beta_{\rm max}$ is already ruled out by binary pulsar observations
\cite{Freire:2012mg}.

\begin{figure}
\center
\includegraphics[width=\columnwidth]{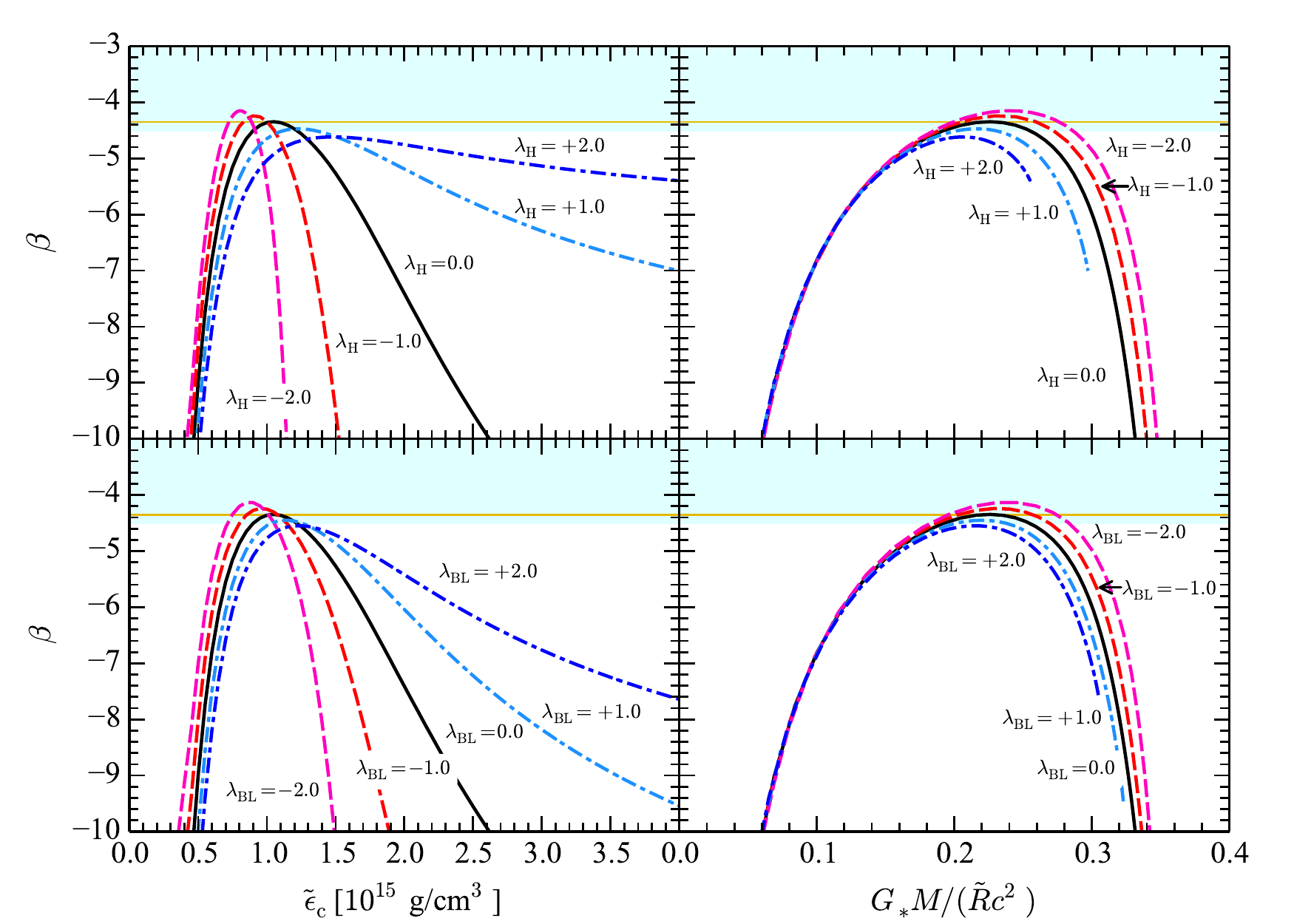}
\caption{Left panels: $\beta$ versus critical central densities
  for different values of $\lambda_{\rm H,BL}$. Right panels:
  $\beta$ versus compactness $G_* M/\tilde{R}c^2$ of the critical
  solutions for different values of $\lambda_{\rm H,BL}$. }
\label{fig:critbeta}
\end{figure}

In the left panels of Figure~\ref{fig:critbeta} we analyze the
dependence of the critical $\beta$ on the central density, focusing on
EoS APR and selecting different values of the anisotropy parameters
$\lambda_{\rm BL}$ (top) and $\lambda_{\rm H}$ (bottom). The shaded
region at the top ($\beta\gtrsim -4.5$) is allowed by current binary
pulsar observations \cite{Freire:2012mg,Wex:2014nva}. The horizontal
line is the roughly EoS-independent threshold $\beta_{\rm max}\simeq
-4.35$ for isotropic stars.
For a given theory, the starting and ending points of the
scalarization regime are those for which a $\beta= {\rm constant}$
(horizontal) line crosses the curves. Anisotropic models have two
distinctive features: (1) when the tangential pressure is larger than
the radial pressure (dashed lines in Figure~\ref{fig:critbeta})
scalarization can occur even for $\beta\geq -4.35$ (for example, for
the Horvat et al. model with $\lambda_{\rm H}=-2$ we have $\beta_{\rm
  crit}= -4.1513$, and for the Bowers-Liang model with $\lambda_{\rm
  BL} = -2$ we have $\beta_{\rm crit} = -4.1354$;
cf. Table~\ref{tab:scalar}, Figure~\ref{fig:ro_scalar_BL} and
Figure~\ref{fig:ro_scalar_QL}); (2) when the tangential pressure is
smaller than the radial pressure (dash-dotted lines in Figure
\ref{fig:critbeta}) scalarized solutions may exist for a much wider
range of $\rhoJ_{\rm c}$.

In the right panels of Figure~\ref{fig:critbeta} we plot the critical
$\beta$ as a function of the stellar compactness
$G_*M/\til{R}c^2$. For low compactness ($M/\til{R}\lesssim 0.15$) all
curves have the same behaviour regardless of $\lambda_{\rm H}$ or
$\lambda_{\rm BL}$. This universality has two reasons: (1) all modern
nuclear-physics based EoS have the same Newtonian limit
(cf.~\cite{Pani:2010vc} for an analytic treatment of this regime for
constant density stars); (2) for any given EoS, the effects of
anisotropy are suppressed in the Newtonian regime, where pressures and
densities are low and the local compactness parameter is small:
cf. Eqs.~\eqref{horvat} and \eqref{BLmod}.

\section{Conclusions}
\label{sec:conclusions}

Binary pulsar observations require $\beta\gtrsim -4.5$
\cite{Freire:2012mg,Wex:2014nva}, and even more stringent constraints
are expected in the near future. As shown in Figure~\ref{cri_bet_eos},
most ``ordinary'' nuclear-physics based EoSs for nuclear matter
predict that scalarization can only occur for $\beta<\beta_{\rm
  max}=-4.35$. As binary pulsar observations get closer and closer to
the limit $\beta\gtrsim -4.35$, the spontaneous scalarization
mechanism originally proposed by Damour and Esposito-Far\`ese
\cite{Damour:1993hw,Damour:1996ke} looks more and more unlikely to be
realized in Nature if neutron stars are isotropic.

The admittedly unlikely event of a binary-pulsar observation of
scalarization with $\beta>-4.35$ would be strong evidence for the
presence of anisotropy, and it may even lead to experimental
constraints on the Skyrme model and QCD. An important caveat here is
that {\em fast} rotation can also strengthen the effects of
scalarization: according to \cite{Doneva:2013qva}, scalarization can
occur for $\beta<-3.9$ when NSs spin at the mass-shedding
limit. However the NSs found in binary pulsar systems are relatively
old, are they are expected to spin well below the mass-shedding limit,
where the slow-rotation approximation works very well
\cite{Berti:2004ny}.

Our work can be extended in several directions. An obvious extension
is to consider the effects of anisotropy at second or higher order in
the Hartle-Thorne expansion. This would allow us to assess whether the
recently discovered ``I-Love-Q'' and ``three-hair'' universal
relations between the multipole moments of the spacetime hold in the
presence of anisotropy {\em and} scalarization
\cite{Yagi:2013bca,Yagi:2013awa,Pappas:2013naa,Yagi:2014bxa,Pani:2014jra}.
A second obvious extension could consider fast rotating, anisotropic
stars (cf.~\cite{Doneva:2013qva,Doneva:2014faa}) and the orbital and
epicyclic frequencies around these objects \cite{Doneva:2014uma}.

Anisotropy can lower the threshold for scalarization to occur, and
this could be of interest to test scalar-tensor theories through
gravitational-wave asteroseismology
\cite{Sotani:2004rq,Sotani:2005qx,Silva:2014ora}.
We also remark that our study used simplified, phenomenological models
for anisotropy, when of course it would be desirable to study
realistic microphysical models. Last but not least, our study should
be extended to evaluate the stellar sensitivities
\cite{Will:1989sk,Zaglauer:1992bp} and to identify exclusion regions
in the $(\beta,\,\lambda)$ parameter space using binary pulsar
observations (cf. e.g.~\cite{Alsing:2011er}).

\section*{Acknowledgments}
We are grateful to K.~Glampedakis, M.~Horbatsch and G.~Pappas for
discussions, and to G.~Pappas for providing EoS data tables.
H.~O.~S. and E.~B. were supported by NSF CAREER Grant No.~PHY-1055103.
C.~F.~B.~M and L.~C.~B.~C would like to thank the Conselho Nacional de
Desenvolvimento Cient\'ifico e Tecnol\'ogico (CNPq), Coordena\c{c}\~ao
de Aperfei\c{c}oamento de Pessoal de N\'ivel Superior (CAPES), and
Funda\c{c}\~ao Amaz\^onia de Amparo a Estudos e Pesquisas do Par\'a (FAPESPA)
for partial financial support. The authors also acknowledge support
from the FP7-PEOPLE-2011-IRSES Grant No.295189.
The figures were created using the
\textsc{python}-based library \textsc{matplotlib} \cite{Hunter:2007}.

\appendix

\section{Derivation of equation (\ref{inertia})}
\label{sec:I}

In this Appendix we present a derivation of the integral
(\ref{inertia}), used to compute the moment of inertia $I$ of slowly
rotating stars in scalar-tensor theory. We begin by noting that
\begin{eqnarray}
\frac{d\Lambda}{dr} = \frac{r}{r - 2\mu}\left(  \frac{1}{r} \frac{d\mu}{dr} - \frac{\mu}{r^2} \right),
\label{dlambda}
\end{eqnarray}
where Eq.~(\ref{lambda}) implies that $\Lambda = - (1/2) \log \left( 1
- 2\mu / r \right)$ and where $d\mu / dr$ is given by
Eq.~(\ref{dmu}). Introducing the auxiliary variable
%
$j \defeq e^{-\Phi - \Lambda}$
%
we find, using Eqs.~(\ref{dphi}) and (\ref{dlambda}), that
\be
\frac{dj}{dr} = - j \, \left[   4\pi A^4(\vp) \frac{r^2}{r - 2\mu} (\rhoJ + \pJ) + r\psi^2  \right].
\ee
Multiplying the frame dragging equation (\ref{domegab}) by $j$ and
rearranging, we obtain
\begin{eqnarray}
\frac{1}{r^4} \frac{d}{dr} \left(  r^4 j \frac{d\omegab}{dr} \right) &= 16 \pi A^4(\vp) \frac{j\, r^2}{r - 2\mu} (\rhoJ + \pJ)  \left( 1 - \frac{\sigmaJ}{\rhoJ + \pJ} \right) \frac{\omegab}{r}.
\end{eqnarray}
If we multiply by $r^4$, integrate from $r=0$ to infinity and use the
fact that
\be
j = 1+{\cal O}(r^{-1}), \quad {\rm{and}} \quad \frac{d\omegab}{dr}  = \frac{6\,  I \, \Omega}{r^4}+{\cal O}(r^{-5}).
\ee
as $r \to \infty$, we finally get Eq.~\eqref{inertia}.

\section*{References}

\providecommand{\newblock}{}

\end{document}